\documentclass[conference, a4paper]{IEEEtran}

\usepackage[utf8]{inputenc}
\usepackage{amssymb}
\usepackage{setspace}
\usepackage{color}
\usepackage[dvipsnames]{xcolor}
\usepackage{graphicx}
\usepackage{breqn}
\usepackage{hyperref}
\usepackage{array}
\usepackage{soul}
\usepackage{lipsum}
\usepackage[utf8]{inputenc}
\usepackage{mathtools}
\usepackage{comment}
\usepackage{epsfig}

\usepackage{cite}
\usepackage{amsmath,amssymb,amsfonts}
\usepackage{lipsum}
\usepackage{multicol}
\usepackage{textcomp}
\usepackage{stfloats}
\usepackage[normalem]{ulem}

\definecolor{boristext}{rgb}{0.22, 0.44, 0.88}
\definecolor{boriscomments}{rgb}{0.88, 0.04, 0.04}
\definecolor{boristochange}{rgb}{0.2, 0.8, 0.8}

\definecolor{rashidtext}{rgb}{0.10, 0.80, 0.20}
\definecolor{rashidcomments}{rgb}{0.4, 0.04, 0.90}

\begin{document}
\title{A Federated Reinforcement Learning Framework for Link Activation in Multi-link Wi-Fi Networks}

\author{\IEEEauthorblockN{Rashid Ali and Boris Bellalta}
\IEEEauthorblockA{Wireless Networking Group / DTIC @ Universitat Pompeu Fabra, Barcelona\\
Email: \{rashid.ali,boris.bellalta\}@upf.edu}}

\maketitle

\begin{abstract}
Next-generation Wi-Fi networks are looking forward to introducing new features like multi-link operation (MLO) to both achieve higher throughput and lower latency. However, given the limited number of available channels, the use of multiple links by a group of contending Basic Service Sets (BSSs) can result in higher interference and channel contention, thus potentially leading to lower performance and reliability. In such a situation, it could be better for all contending BSSs to use less links if that contributes to reduce channel access contention. Recently, reinforcement learning (RL) has proven its potential for optimizing resource allocation in wireless networks. However, the independent operation of each wireless network makes difficult --- if not almost impossible--- for each individual network to learn a good configuration. To solve this issue, in this paper, we propose the use of a Federated Reinforcement Learning (FRL) framework, i.e., a collaborative machine learning approach to train models across multiple distributed agents without exchanging data, to collaboratively learn the the best MLO-Link Allocation (LA) strategy by a group of neighboring BSSs. The simulation results show that the FRL-based decentralized MLO-LA strategy achieves a better throughput fairness, and so a higher reliability ---because it allows the different BSSs to find a link allocation strategy which maximizes the minimum achieved data rate--- compared to fixed, random and RL-based MLO-LA schemes. 
\end{abstract}

\IEEEpeerreviewmaketitle
\section{Introduction}

Several new applications requiring reliable high-throughput and low-latency are proliferating, such as augmented and virtual reality (AR/VR), online 3D gaming, holographic teleconferencing, and cloud computing \cite{GalGerCar2023,adame2021time}. 
In the search to meet these new needs, IEEE 802.11be~\cite{Ref01} ({next-generation Wi-Fi 7}) has introduced a new feature called multi-link operation (MLO) capability \cite{marcUnderstandingMLO}. It enables multiple concurrent links ---each working on a different channel--- between an access point (AP) and its associated stations (STA). It is expected MLO will further developed in Wi-Fi 8 by combining it with Multi-AP cooperation strategies \cite{GalGerCar2023}.

\begin{figure}[t]
    \centering
    \includegraphics[width=\columnwidth]{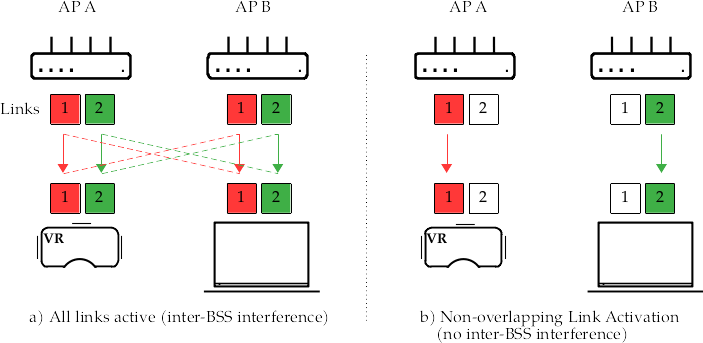}
    \caption{A Link Activation mechanism can improve the system performance by reducing mutual interference between MLO BSSs.}
    \label{Fig:EXampleFig}
\end{figure}

The use of multiple-links per basic service set (BSS) increases the chances to overlap with other neighboring BSSs since the number of available channels remains the same, and so increasing inter-BSS interference. Thus, in scenarios with multiple overlapping BSSs (OBSS) it has been observed that the use of MLO can be counterproductive \cite{marcUnderstandingMLO}, as it may increase network delay and decrease network throughput \cite{marc80211beMLO}. This situation is illustrated in Fig. \ref{Fig:EXampleFig} where two neighboring multi-link BSSs share the same channels in both links. In the left-side (case a) both BSS simultaneously transmit using the two available links, thus suffering from mutual interference. In the right-side, each BSS decides to use a single but different link, resulting in an interference free transmission. Then, even when multiple links are available, it could be better in terms of performance to only use a subset of them. Therefore, the decision of which links need to be active and not is of crucial importance.

In this paper, we show that empowering APs with machine learning (ML) agents helps to improve their ability to activate a proper subset of links to use. 
The use of ML has proven its strengths to tackle many of the challenges related to resource allocation in Wi-Fi networks \cite{WiFiML}. While the use of supervised and unsupervised ML techniques are not suitable for the most of the problems related to the needs and challenges in uncertain growing Wi-Fi networks \cite{mabwifi}, mainly due to the need to have trained datasets, Reinforcement Learning (RL) emerged as the most favorable ML technique to tackle dynamic resource allocation problems because it learns the model by interacting with the environment and adapts to particular scenarios \cite{RLWiFi01,RLWiFi02}. 

In an RL-based framework, an AP as an agent exploits its best available actions based on the highest accumulated reward as a feedback of the selected action. However, in dense scenarios with multiple overlapping BSSs interfering each other over multiple links, an RL-based solution may not be effective due to the complex inter-BSS dependencies, and the adversarial-like decisions of multi-agent RL (MARL) environments. A MARL setup can have several challenges \cite{MARL} for a decentralized OBSSs due to the increased complexity and dynamics of the environment. Such as, in MARL, each agent's behavior affects the environment, which can change rapidly and unpredictably. This non-stationarity makes it difficult for each individual agent to learn and adapt to the environment. Similarly, it is also challenging to assign a reward to each agent's actions taking into account the overall performance of the environment. Since the actions of one agent can affect the rewards of other agents, it can be difficult to determine which actions were responsible for the resulting performance. Another critical problem that may arise in a multi-agent RL environment is the curse of dimensionality: as the number of agents increases, the number of possible joint actions and states increases exponentially, making the learning problem more challenging. 

To solve this situation, we consider a collaborative approach to learn the best action for each BSS, and leverage the use of Federated Reinforcement Learning (FRL) \cite{FederatedRL}. FRL is a collaborative and distributed ML technique, which enables multiple agents to create a common and robust learning model without sharing the actual data. In a distributed FRL-based framework, each local agent shares their local model to all other agents, which in turn compute the global learning model from the received ones \cite{WiFiFRL}.

The main contributions of this paper are summarized as follows:
\begin{itemize}
    \item We describe the extra degrees of freedom introduced by multi-link wireless networks, and how MLO can improve data rates in uncoordinated dense deployments by properly activating only a subset of links in each BSS.
    \item To learn which is a feasible subset of links in each BSS, we introduce a Federated Reinforcement Learning framework, where all neighboring APs share their instantaneous performance (i.e., reward), and so all of them are aware of the impact of their actions on the neighborhood.    
    \item We evaluate the introduced FRL framework by simulation in multiple different scenarios. The presented results confirm that the proposed FRL framework notably outperforms both the case where all BSSs have all links always active, and the case where only local learning is used. 
    \item Our proposed FRL-based scheme is designed to improve fairness and reliability in a wireless environment. This is achieved by maximizing the minimum data rate achieved by each BSS in the network. By doing so, it ensures that all BSSs receive a fair and reliable allocation of network resources.
    %
\end{itemize}

\section{System Model}\label{sec:systemModel}


\subsection{Scenario}

We consider a scenario consisting of a set $\mathcal{A}$ of $n$ APs, that is $\mathcal{A} = \{A_{1},A_{2}...,A_{n}\}$, deployed over a certain area. Each AP has a single STA associated to it. All AP-STA pairs use the same set of $k$ available links $\mathcal{L}=\{L_{1},L_{2}...,L_{k}\}$ for communication, and so they interfer each other when placed nearby. Only downlink communication is considered. 

An instance of the system model is evaluated for $T$ iterations, with $t \in \{1, 2, 3 ..., T\}$, the temporal index. At iteration $t$, each AP can decide to activate all links in $\mathcal{L}$ or only a subset of them following a particular MLO-LA strategy. 

\subsection{Pathloss Model}

The distance between each AP and its associated STA is the same in all BSSs and equal to $d$ meters. 
To compute the pathloss between them on each link, we use the TMB pathloss model proposed in \cite{pathloss} for $5$~GHz IEEE 802.11ax residential scenario, and is given by,
\begin{equation}
   \text{PL}\left(d\right) = L_{0} + 10 \times \gamma \times \log_{10}\left(d\right) + c \times \overline{W} \times d 
\end{equation}\label{eq:pathloss}
where $L_{0}$ is the pathloss intercept and $\gamma$ is the attenuation factor. The factors $c$ and $\overline{W}$ are used as the attenuation of each wall and the average number of traversed walls per meter, respectively.
\subsection{Achievable Transmission Rate}

At time $t \in \{1, 2, 3 ..., T\}$, the data rate achieved by AP $i$ is given by
\begin{equation}
R_{i,t} = \sum_{\forall L_k \in \mathcal{L}}{\alpha_{L_k} B_{i,L_k,t} \times \log_{2}\left(1 + \frac{P_{i,t}}{I_{i,L_k,t} + N}\right)},
\end{equation}
where $B_{i,L_k,t}$ is the bandwidth of the link $L_k$, $P_{i,t}$ is the received power at its associated station, determined by \eqref{eq:pathloss} and the transmission power, and $I_{i,L_k,t}$ and $N$ are the sum of the interference observed by the station associated to AP $i$ in link $L_k$, and the floor noise power, respectively. Note that $I_{i,L_k,t}$ is calculated as the aggregate power received at the station associated to AP $i$ from all other APs that are using the link $L_k$. $\alpha_{L_k}$ is 1 if the link is active, and 0 otherwise.


\subsubsection*{Multi-link Operation Link Adaptation strategies}

We consider the following four MLO-LA strategies:
\begin{enumerate}
    \item Fixed LA scheme: All APs have all their links activated all the time.
    \item Random LA scheme: Each AP activates one or more links randomly at every iteration. 
    \item RL-based LA scheme: Each AP uses a Multi-Armed Bandit (MAB) algorithm (e.g., $\varepsilon$-greedy) to learn which is the most suitable link activation strategy to implement. It exploits the best performing configuration by selecting the one with the highest accumulated average reward. {It updates the set of activated links at every iteration same as in the previous case.} 
    \item FRL-based LA scheme: Each AP receives the rewards from neighboring APs and selects the minimum reward to rate the performance of its last selected link activation strategy. Link activation strategies at each node are selected using $\varepsilon$-greedy algorithm as in the previous case.
\end{enumerate}

\section{FRL framework for decentralized link activation} \label{Sec:Learning}

To build our FRL-based framework we start by defining the actions set $\mathcal{O}_{i}$ for AP $i$, which includes all the possible LA combinations, and is given by $\mathcal{O}_{i} = \{O_{1},O_{2},…,O_{p}\}$, with $p$ the total number of possible combinations based on the number of available links. It is given by $p = 2^{k}-1$, with $k$ the number of links. Note we do not include as a possible option the case where an AP does not select any channel for transmission.

The FRL-based framework is composed of two learning models: a) a local learning model (LLM), and b) a global learning model (GLM). In the LLM, each AP uses a RL-based learning solution, such as MABs, that only relies on the individual AP performance to rate current configuration. Despite the excellent performance that RL-based solutions have achieved in decentralized spatial reuse and spectrum allocation techniques \cite{wilhelmi2019potential,barrachina2021multi}, they suffer in multi-agent environments because each individual agent is taken actions selfishly, thus preventing in most of the cases to reach a satisfactory equilibrium for all of them. Therefore, multi-agent environments can be significantly improved by allowing them to collaborate \cite{wilhelmi2019collaborative}. To do so, our FRL framework implements a GLM where AP shares with its neighbouring APs its LLM reward as a local experience, from where the GLM reward is computed to establish a global learning. 
\subsection{Local Learning Model}

In our problem, we have a limited number of radio links shared among all AP-STA pairs. These limited number of resources must be allocated among competing APs in a way that maximizes their expected gain. {Since each APs' selections are only partially known at the time of allocation, we formulate our local learning model as a MAB problem, where each AP has $p$-arms (as described earlier, $p$ is the total number of possible choices of the available radio links).} A MAB problem models an interaction between a learning agent, and the environment (e.g., set of AP-STA pairs). The agent can choose one action every round $t = \{1, 2, 3, ...,T\}$ from a set of actions $\mathcal{O}$. For each arm pulled, $o$ the learning agent placed at AP $i$ receives an instant reward ($\Omega_{i}^{o}(t)$) from the environment, which is used to evaluate the performance of the selected action, as well as to select subsequent actions. The goal of the learning agent is to maximize the long-term reward to find an optimal action ($o^{*}$). Thus, every time an AP $i$ as a learning agent activates a certain set of link(s) following action $o$, it obtains an achieved data rate as its instantaneous reward
\begin{equation}
\Omega_{i,t}^{o} = R_{i,t}^{o}.
\end{equation}\label{eq:InstReward}

In this way, each of the actions (i.e., set of  activated links) is instantly rewarded as a function of the observed transmission rate. Then, AP $i$ accumulates its instant reward $\Omega_{i,t}^{o}$ for a given action $o$ at time $t$ to compute the average reward $\widehat{\Omega_{i,t}^{o}}$, given by,
\begin{equation}
\widehat{\Omega_{i,t}^{o}} = \frac{1}{V^{\prime}} \sum_{v=1}^{V\prime} \Omega_{i,v}^{o}.
\end{equation}\label{eq:AccuReward}
where $V\prime$ is the number of times an action $o$ has been selected by the AP $i$ until iteration $t$. Note that by using the achieved transmission rate $R_{i,t}$ as a reward metric for the LLM, we intend to make each AP capable to keep track of the network interference, i.e., high interference results in a lower transmission rate, and so in a low reward. 


The strategy to ensure a good network performance from each APs’ perspective is to avoid channel overlapping with nearby APs, even if that means to use only a subset of the available links. Therefore, at time $t$, the agent aims to activate the set of link(s) with the highest accumulated average reward (i.e., the optimal action from the perspective of the agent), that is, 
\begin{equation}
    o_{t}^{*} = \textit{arg}\max_{\forall o \in O}{\widehat{\Omega_{i,t}^{o}}}.
\end{equation}

A MAB algorithm involves the exploration and exploitation trade-off, in which the agent must deal between learning at a faster or slower pace. To manage this trade-off, a learning rate parameter ($\varepsilon$) is used to balance both exploration with probability $\varepsilon$ and exploitation with probability $1-\varepsilon$. Note that a faster learning rate may lead to not exploring enough, ending into a sub-optimal solution, whereas a slower learning rate may waste too much time on bad decisions. Therefore, tuning and selecting the appropriate learning rate is fundamental in order to achieve good results. 

\subsection{Global Learning Model}

To solve the adversarial-like decisions taken by individual agents in the LLM, we propose to use a global learning model on the top of the LLM to collaboratively optimize the resource allocation in terms of link(s) activation. In our GLM, each AP shares its locally computed reward to its neighbouring APs using periodic beacon-like messages. To implement the GLM, each AP uses its minimum achieved reward (MAR) function as its instant global reward, which selects minimum reward from the neighbours. Therefore, the reward of an AP $i$ at time step $t$ for an action $o$ is the minimum reward received by an AP among the contributing neighbours, that is,
\begin{equation}\label{eq:LLM}
\text{MAR}^{o}(i,t) = \arg \min_{\forall n^{\prime} \subseteq n}{\widehat{\Omega_{n^{\prime},t}^{o}}}
\end{equation}\label{eq:InsGLM}

We use max-min fairness strategy \cite{minMaxFRL}, where an FRL agent (AP $i$) collects all the rewards ($\Omega^{o}$) from the neighboring APs $n^{\prime}$ ($n^{\prime} \subseteq n$) and selects the minimum reward as its instant global reward, $\rho_{i,t}$.

The concept of max-min fairness is a widely recognized method for establishing fairness in network resource allocation \cite{minMaxFRL}. This method aims to maximize the minimum achievable data rate for all participating users. In our scenario, we have a wireless network that comprises links with fixed capacities and a set of AP-STA pairs that communicate over these links.

Finally, each AP $i$ computes the mean global reward for a given action $o$, 
\begin{equation}\label{eq:AccuGLM}
\widehat{\rho_{i,t}^{o}} = \frac{1}{V^{\prime}} \sum_{v=1}^{V^{\prime}}{\rho^{o}_{i}(V^{\prime})},    
\end{equation}
which is used to select next action in exploitation mode (i.e., with probability $1-\varepsilon$) as the agent aims to maximize it, i.e., $o^*=\arg \max_{\forall o \in O}{\widehat{\rho_{i,t}^{o}}}$.


\section{Performance Evaluation}\label{sec:performanceEvaluation}


\begin{figure}
                \centering
                \includegraphics[width=0.8\columnwidth]{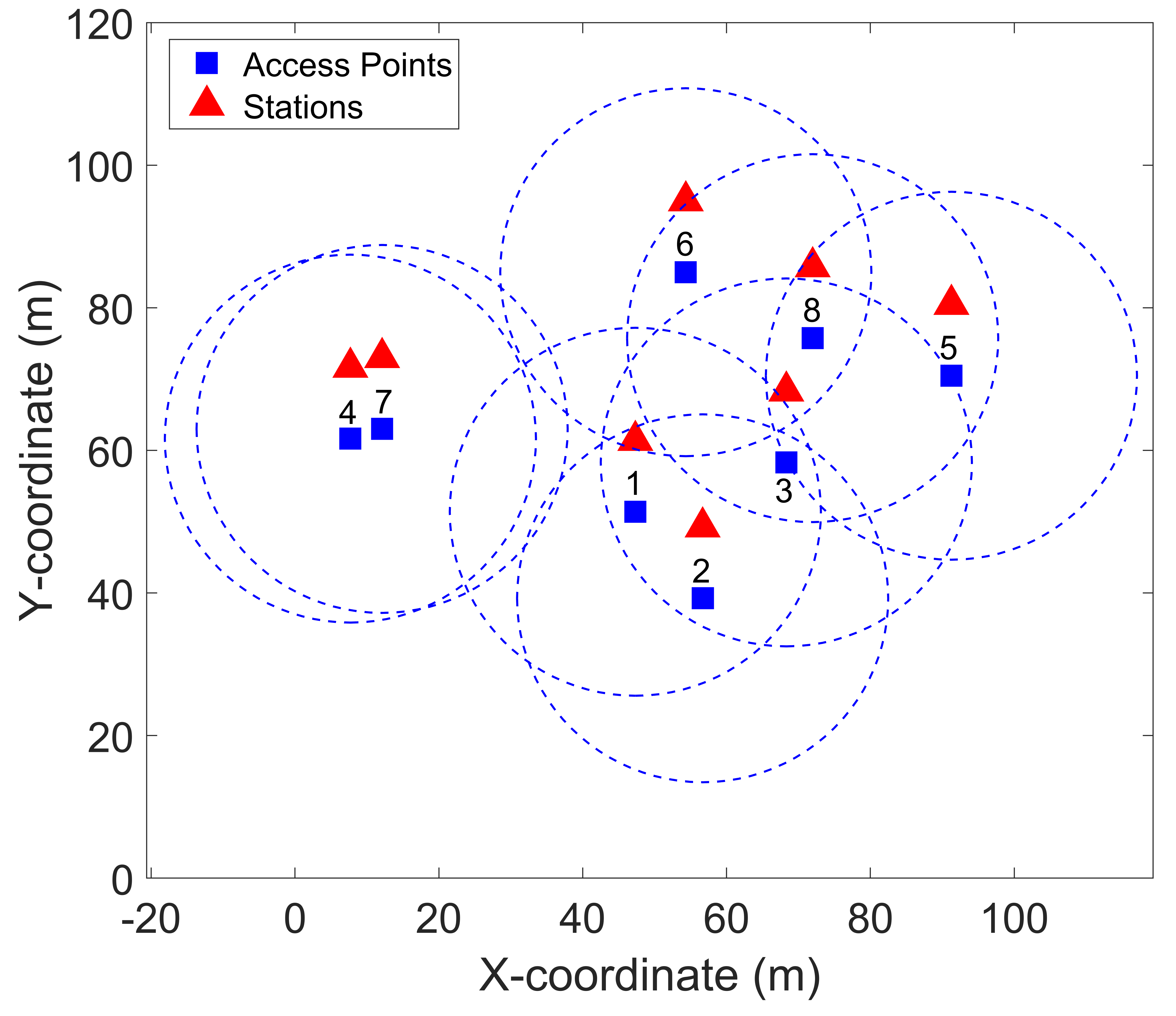}
                \caption{A scenario of randomly placed eight APs (each associated to a single station using four radio links) in a $100\times100$ meters area. {Note the blue dashed circles around the APs show network coverage of an AP taking into account a sensitivity of -82 dBm.}}
                \label{fig:fig02}
\end{figure}
\subsection{Simulation set-up}

\subsubsection*{Scenario}
We evaluate the performance of our proposed MLO-LA FRL-based framework in a scenario consisting of a group of eight randomly placed APs in an area of $100\times100$ meters. Figure \ref{fig:fig02} depicts a particular scenario with eight APs, where each AP has a single station associated. The station is deployed at a distance of 10 meters from its AP. All APs and stations are equipped with four radio interfaces. We only consider downlink traffic. 


\subsubsection*{Parameters}

In our simulations, we consider $500$ different scenarios, with each scenario lasting for $2000$ iterations. The simulation results are obtained using the TMB path-loss parameters \cite{pathloss} proposed for IEEE 802.11ax \cite{SimulationScenario}: $L_{0} = 54.12$, $\gamma = 2.06067$, $c = 5.25$, and $\overline{W} = 0.1467$. The data rates are obtained using a transmission power of $20$ dBm and $80$ MHz channel bandwidth per link. Sensitivity threshold for reception is set to $-82$ dBm. With respect to $\varepsilon$-greedy algorithm, a slow-decreasing $\varepsilon$ value with respect to the number of iterations (that is, $\varepsilon = \frac{1}{\sqrt{t}}$) is used, which allows an agent to explore the environment at the beginning and exploit it afterwards. 


In our proposed FRL-based MLO-LA scheme each AP shares its locally computed reward, i.e., $\Omega_{i,t}^{o}$ from~\eqref{eq:InstReward}, with the APs within its coverage area, and computes the global reward, i.e., $\text{MAR}^{o}(i,t)$ from~\eqref{eq:InsGLM}, with the received values from the other APs. The proposed FRL-based mechanism allows the APs to choose an optimal action ($o^*$) based on the maximized global expected reward, as in \eqref{eq:AccuGLM}. APs can share with their neighboring BSSs their locally learned reward using the 'beacon' frames to compute the global learning parameters. Note that global in this case means between a group of overlapping APs.

\subsection{Results}


\subsubsection{Single scenario}

\begin{figure}
                \centering
                \includegraphics[width=0.8\columnwidth]{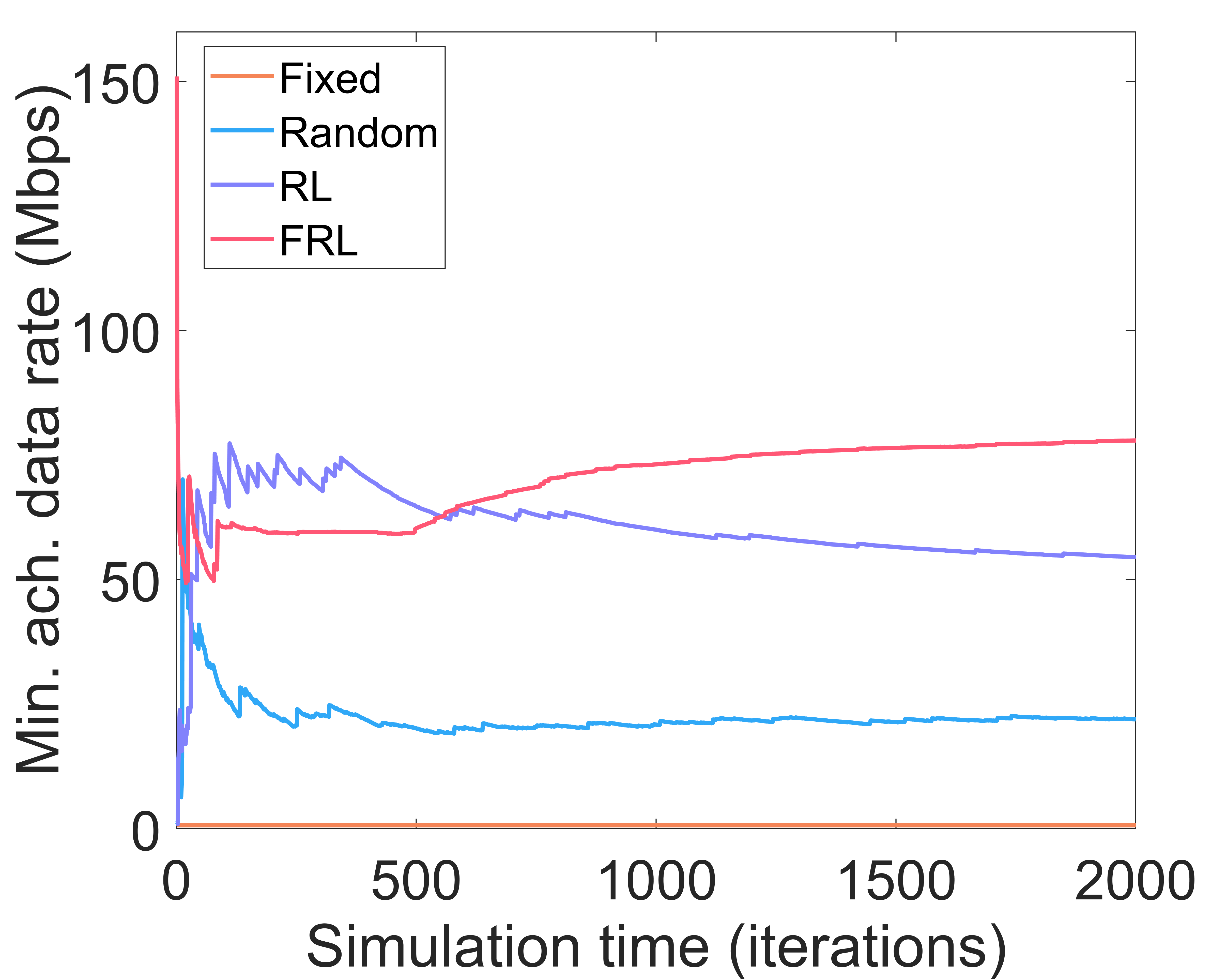}
                \caption{A comparison of temporal evolution of the achieved data rates of the AP with minimum average data rate for fixed, random, RL-based, and FRL-based decentralized link(s) activation schemes in the scenario depicted in Fig. \ref{fig:fig02}.}
                \label{fig:fig03}
\end{figure}

We first consider the scenario depicted in Fig. \ref{fig:fig02}, where we can find different cases of interest. For instance, AP 4 and 7 only overlap with each other, while AP~6 experiences less interference. On the other hand, AP 1, 2, 3, and 8 experience significant interference.
Fig. \ref{fig:fig03} shows the temporal evolution of the AP with minimum average achieved data rate over 2000 iterations. In Fig. \ref{fig:fig03}, we see that for a single scenario where most of the neighbouring BSSs overlap, the fixed LA scheme severely degrades the performance of the entire network. The random LA scheme may work efficiently due to the equal probability of activating any set of links for every BSS, which allows APs to periodically enjoy situations of low interference. Similarly, the RL-based scheme finds the set of link(s) to activate based on locally accumulated reward, and thus performs better than fixed and random LA schemes. In a RL-based MLO-LA scheme, an AP learns to make decisions based on the feedback received from the environment in the form of achieved data rate. On the other hand, fixed and random schemes make decisions without considering any feedback from the environment. This increase in the performance of RL-based MLO-LA scheme shows the importance of the use of ML for link(s) adaptation in MLO. However, since our scenario is a multi-agent environment, where each individual AP ---as an individual agent--- is exploiting selfishly, the achieved performance is sub-optimal. As shown in the figure, the FRL-based MLO-LA scheme converges at higher minimum data rate than the RL-based scheme, which is mainly due to the collaborative exploitation of the available links.

\begin{figure}
                \centering
                \includegraphics[width=0.8\columnwidth]{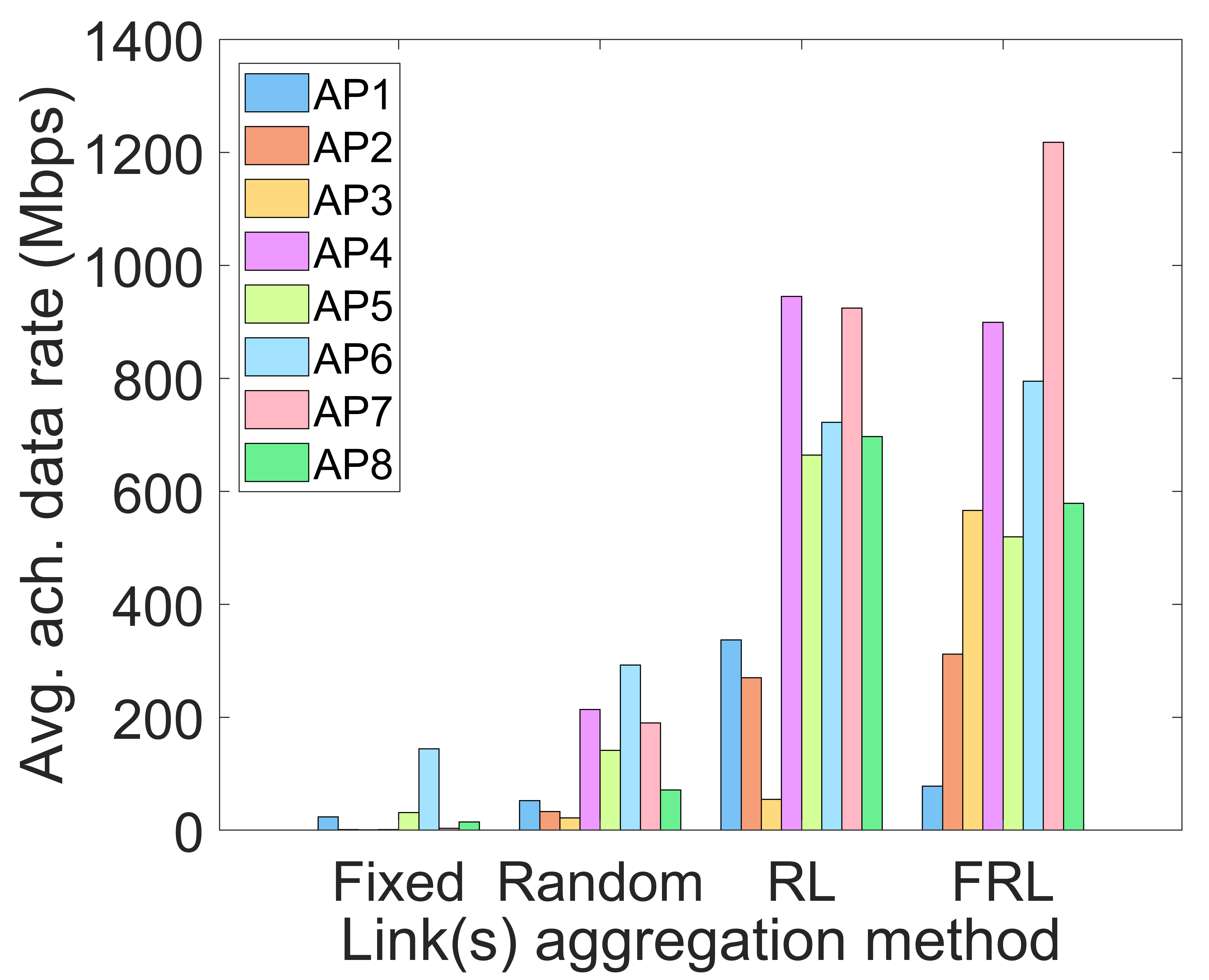}
                \caption{A comparison of achieved data rates averaged over 2000 iterations for fixed, random, RL-based, and FRL-based decentralized link(s) activation schemes in the scenario depicted in Fig. \ref{fig:fig02}.}
                \label{fig:fig04}
\end{figure}

Fig. \ref{fig:fig04} depicts the average achieved data rate for all 8 APs. We observe that in the case of AP 4 and 7, which only overlap with each other, ML-enabled MLO-LA schemes (i.e., RL and FRL) allow both access points to learn a good set of transmission links,
%
%
resulting in higher data rates compared to the fixed and random schemes. Similarly, in the case of AP 6, which experiences the least interference, a significant increase in the achieved data rate is also observed for the FRL-based scheme.



\subsubsection{Multiple scenarios}
We consider now the results over multiple (i.e., $500$) random generated scenarios. 

\begin{figure}
                \centering
                \includegraphics[width=0.8\columnwidth]{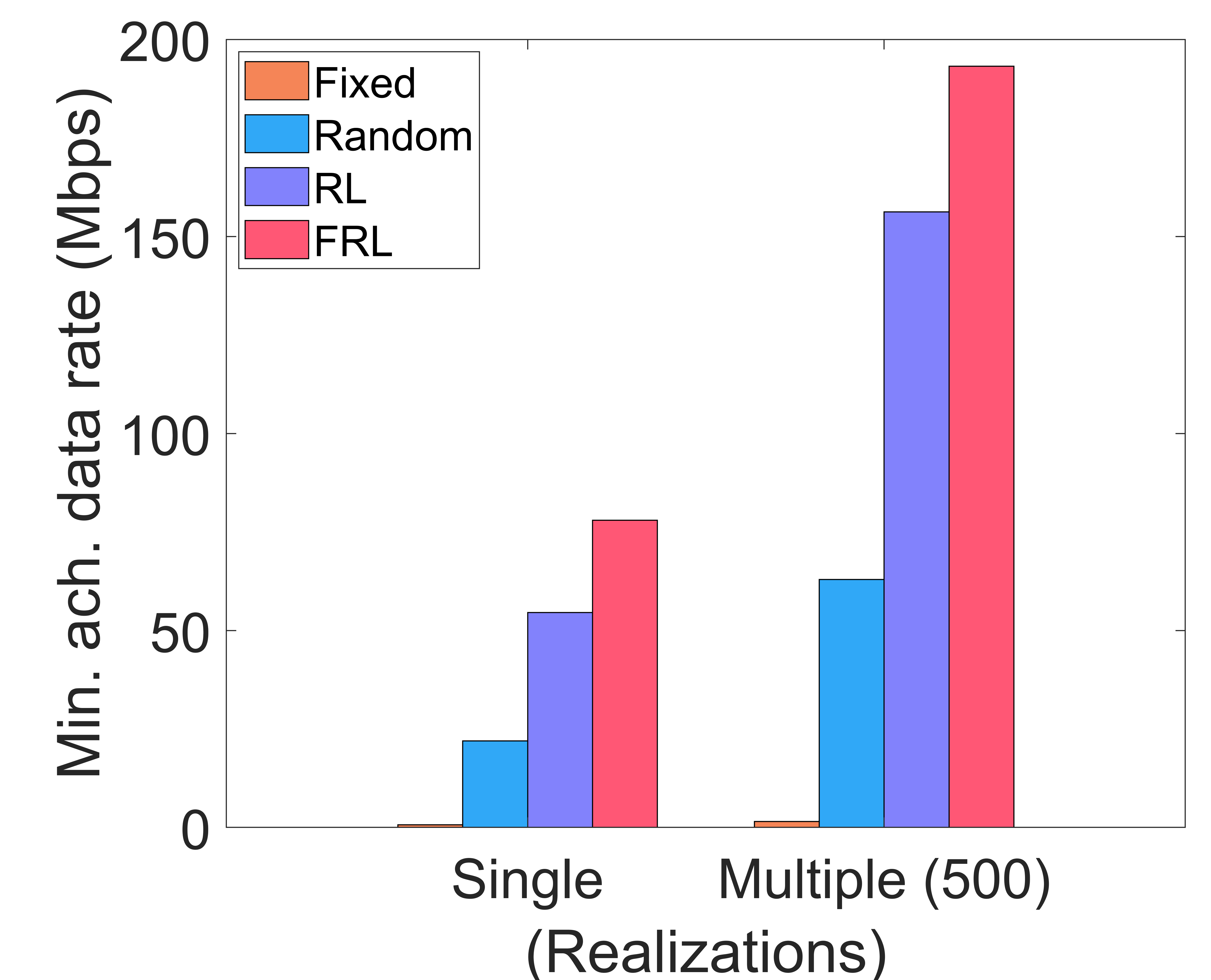}
                \caption{A comparison of average minimum achieved data rate for fixed, random, RL-based, and FRL-based decentralized link(s) activation schemes in eight APs network of; single scenario and multiple ($500$) scenarios (Minimum from the average of $2000$ iterations over the $500$ realizations).}
                \label{fig:fig05}
\end{figure}

Fig. \ref{fig:fig05} shows the average minimum achieved data rate for the fixed, random, RL-based and FRL-based schemes in both the scenario depicted in Fig. \ref{fig:fig02}, and the average over the 500 random scenarios. We observe that in the case of multiple scenarios, FRL offers a significantly higher minimum data rate ($193.212$ Mbps) than random ($62.952$ Mbps) and fixed ($1.485$ Mbps) schemes, respectively. The FRL-based scheme benefits from the use of collaborative network configuration inference and provides an average data rate that is about $40$ Mbps higher than that of the non-collaborative RL-based scheme ($156.204$ Mbps). These results confirm that a collaborative approach to the network resource allocation in a decentralized network environment, as it is the multiple links selection, is able to increase the achievable data rates.

\begin{figure}
                \centering
                \includegraphics[width=0.84\columnwidth]{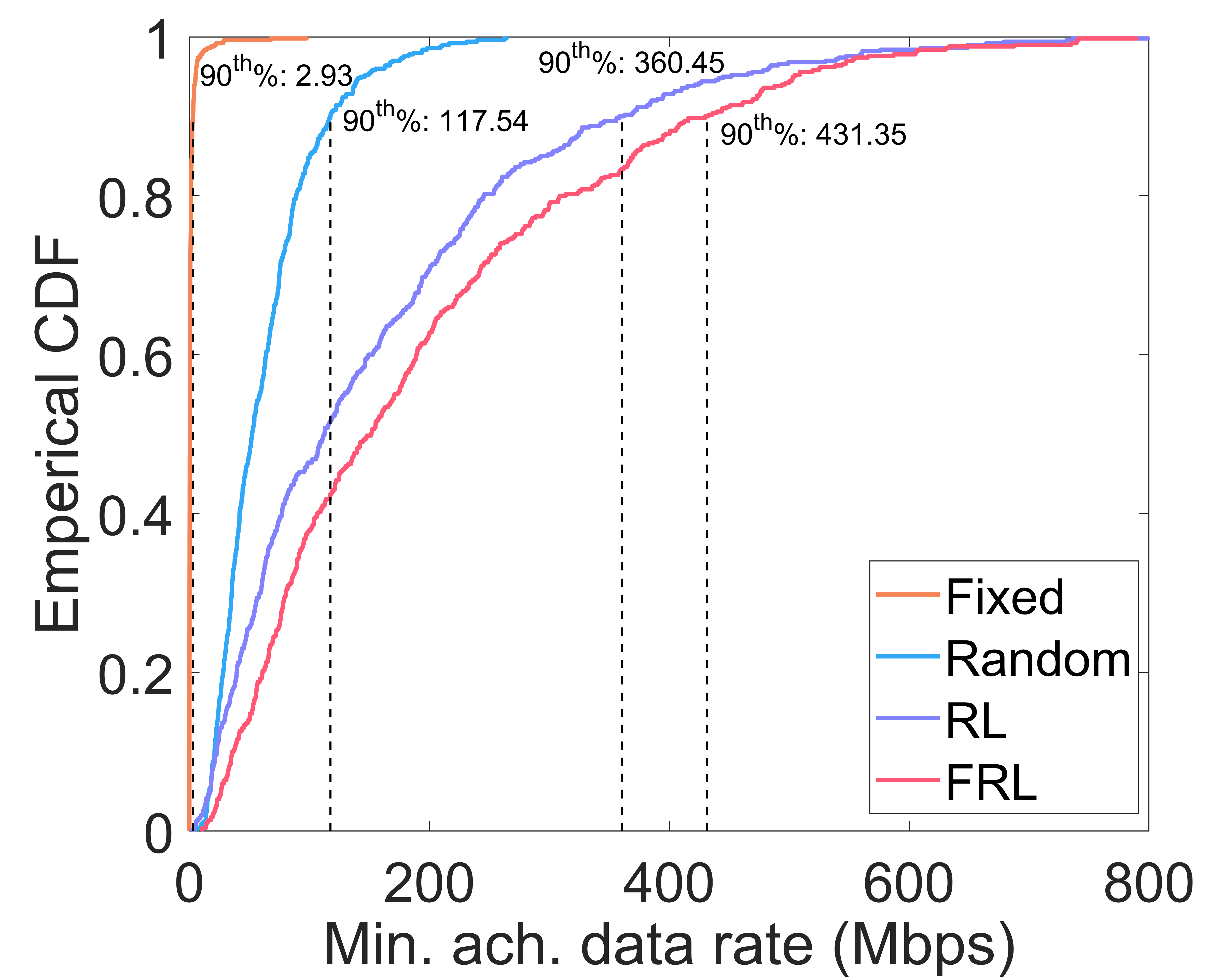}
                \caption{Empirical CDF of the minimum achieved data rates for the four considered LA schemes over $500$ sample scenarios (dashed lines show 90\%-tile for each of the LA scheme).}
                \label{fig:fig06}
\end{figure}

Fig. \ref{fig:fig06} shows the empirical cumulative distribution function (ECDF) of the minimum achieved data rate by each MLO-LA scheme over the $500$ sample scenarios. The minimum achieved data rate of each individual scenario is computed as the average temporal data rate of the AP with the minimum achieved data rate. We observe that activating link(s) using a random scheme may allow APs to achieve higher data rates than using the fixed scheme. However, since the random MLO-LA scheme blindly activates a different set of link(s) at every iteration, it does not learn neither from the previously good or bad configurations tested. Concerning the ML-enabled models, the proposed FRL-based scheme is the one providing the highest minimum achieved data rate (Mbps). The ECDF of the FRL-based scheme shows that in most of the scenarios the minimum achieved data rates are higher thanks to choosing the link(s) based on the information from the interfering APs (i.e., using the lowest reward from the neighbours). Moreover, we can make some general observations based on the $90^{th}$ percentile values of their achieved data rates, as shown in Fig. \ref{fig:fig06}. First, it appears that the FRL-based and RL-based schemes are able to achieve much higher data rates than the fixed and random MLO-LA schemes. This is likely due to their ability to adapt to changing network conditions and learn a proper set of links to activate. Second, there is a significant difference in performance between the FRL-based and RL-based schemes, with the FRL-based scheme achieving a much higher $90^{th}$ percentile value. This shows that the collaborative approach of the FRL-based scheme provides a significant advantage over the non-collaborative RL-based approach.

\begin{figure}
                \centering
                \includegraphics[width=0.8\columnwidth]{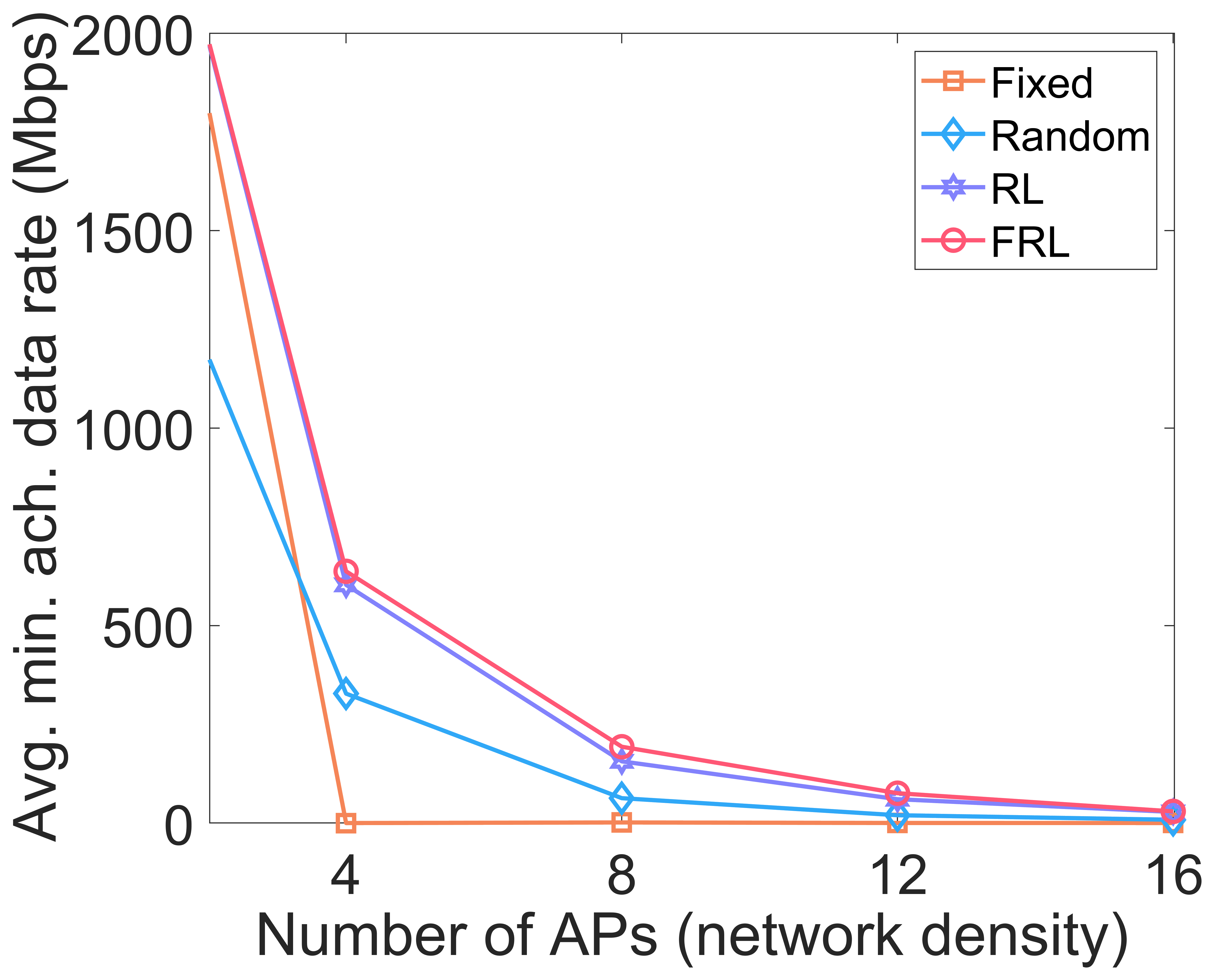}
                \caption{Performance evaluation of the proposed FRL-based link(s) activation scheme in terms of average minimum achieved data rate over all APs for different network densities (i.e., $2, 4, 8, 12$ and $16$).}
                \label{fig:fig07}
\end{figure}

Finally, we provide more insights on the generalization capabilities of the proposed FRL-based scheme for different network densities. Fig. \ref{fig:fig07} presents the minimum data rate achieved by each of the four schemes for different network densities (i.e., for $2$, $4$, $8$, $12$, and $16$ APs). As expected, all MLO-LA schemes perform poorer as the network density increases. However, we can confirm that the FRL-based scheme scales better than the others as the network density increases. 


\section{Conclusions}\label{sec:conclusions}

This work explores the use of an FRL framework for decentralized link(s) activation in multi-link Wi-Fi networks. The simulations in this study compare; fixed use of the available links, random activation of the link(s), RL-based link(s) activation, and FRL-based link(s) activation for the transmission. From the results, we draw following key insights,
\begin{itemize}
    \item Given the ability to exchange data among multiple BSSs through beacons, we find that a collaborative selection of the available link(s) maximizes the minimum achieved data rate--- compared to fixed, random and RL-based MLO-LA schemes.
    \item An FRL-based LA scheme helps to enhance the network performance faster in terms of data rate as compared to the RL-based LA scheme, mainly due to cooperative learning approach.
    \item In sparse to dense network scenarios, that is from few to several randomly deployed APs in a $100~m^{2}$, uncoordinated decentralized LA schemes perform poorer as the network density increases. Whereas, FRL allows selecting the most suitable link(s) to activate, resulting in a higher network performance. 
\end{itemize}

In the present work, we explore the potentials of FRL-based LA scheme for decentralized link(s) activation in multi-link Wi-Fi APs. This work can be extended by investigating the effect on the network performance ---and the response of the different ML algorithms--- of different strategies to calculate the collaborative reward.
\section*{Acknowledgment}
R. Ali has received funding from the European Union's Horizon 2020 research and innovation programme under the Marie Sk\l{}odowska-Curie grant agreement n$^{\circ}~945380$. B. Bellalta and R. Ali are partially funded by Wi-XR PID2021-123995NB-I00 (MCIU/AEI/FEDER,UE).

\bibliographystyle{IEEEtran} 
\bibliography{RA-BB-MLO-FRL-arXiv}

\end{document}